\documentclass[aps,prc,preprint,groupedaddress]{revtex4}

\usepackage{multirow}
\usepackage{epsfig}
\usepackage{textcomp}
\usepackage{youngtab,color}
\usepackage{multirow}
\usepackage{epsfig}
\usepackage{longtable}
\begin{document}

\title{The observed $\Omega_{c}^{0}$ resonances as pentaquark states}
\author{C. S. An}\email{ancs@swu.edu.cn}
\author{H. Chen}\email{chenh@swu.edu.cn}
\affiliation{School of Physical Science and Technology, Southwest University,
              Chongqing 400715, People's Republic of China }

\thispagestyle{empty}

\date{\today}

\begin{abstract}

In present work, we investigate the spectrum of several low-lying $sscq\bar{q}$
pentaquark configurations employing the constituent quark model, within which
the hyperfine interaction between quarks is taken to be mediated by
Goldstone boson exchange. Our numerical results show that four $sscq\bar{q}$
configurations with $J^{P}=1/2^{-}$ or $J^{P}=3/2^{-}$ lie at energies
very close to the recently observed five
$\Omega_{c}^{0}$ states by LHCb collaboration, this indicates that
the $sscq\bar{q}$ pentaquark configurations may form sizable components
of the observed $\Omega_{c}^{0}$ resonances.

\end{abstract}

\maketitle

\section{introduction}
\label{sec:intro}

Recently, LHCb collaboration announced the observation of
 five $\Omega_{c}^{0}$ states named $\Omega_{c}(3000)^{0}$, $\Omega_{c}(3050)^{0}$,
$\Omega_{c}(3066)^{0}$, $\Omega_{c}(3090)^{0}$ and $\Omega_{c}(3119)^{0}$ corresponding to their masses,
by studying the $\Xi_{c}^{+}K^{-}$
mass spectrum with a sample of $pp$ collision data~\cite{Aaij:2017nav}.
As we know, the experimental data about $\Omega_{c}^{0}$ baryons is still
very poor, only the two ground states have been found before the LHCb
observation~\cite{pdg}. Consequently, observation of the five states
is of significant importance for us to understand structure of
the $\Omega_{c}^{0}$ baryons.

Seeing the LHCb experimental results, one may immediately ask
why there are five $\Omega^{0}_{c}$ states with narrow width
lying at $\sim3000$~MeV, and what kind of structure of these
states should be. In fact,
just after the LHCb announcement, various of theoretical investigations on the structure and quantum numbers of $\Omega^{0}_{c}$
resonances have been done, using the
QCD sum rule method~\cite{Agaev:2017jyt,Chen:2017sci,Wang:2017zjw,Aliev:2017led},
quark-diquark picture~\cite{Karliner:2017kfm,Wang:2017vnc,Cheng:2017ove},
chiral quark model~\cite{Yang:2017rpg,Huang:2017dwn}, constituent quark model~\cite{Wang:2017hej},
lattice QCD~\cite{Padmanath:2017lng}, chiral quark-soliton model~\cite{Kim:2017jpx},
and in several very recent works, the decay behaviours of the observed
$\Omega^{0}_{c}$ are also investigated~\cite{Zhao:2017fov,Chen:2017gnu,Agaev:2017lip}.

On the other hand, one may notice that two hidden charm pentaquark states
have been announced by LHCb collaboration in 2015~\cite{Aaij:2015tga}, one
of these two states also has narrow decay width.
And in a recent work, the spectrum of low-lying $sssq\bar{q}$
configurations has been investigated~\cite{Yuan:2012zs}, the results show us that the lowest pentaquark
configuration with $S=-3$ and negative parity lies at lower energy
than the lowest $sss$ three-quark resonance predicted by traditional
quark model. In another word, baryon excitation by pulling out a $q\bar{q}$ pair may
need less energy than that by the traditional orbital excitation. Since there are at least
two strange quarks and one charm quark in the $\Omega^{0}_{c}$ resonance, the
pentaquark configurations with light $q\bar{q}$ must play very special roles in the
properties of these resonances. Therefore,
in present work, we study the spectrum of several low-lying $sscq\bar{q}$
($q\bar{q}$ denotes a light quark-antiquark pair with isospin $I=0$) configurations
with negative parity employing the constituent quark model, within which
the hyperfine interaction between quarks are taken to be mediated by
Goldstone boson exchange, to investigate whether the pentaquark
configurations can form sizable components in the observed $\Omega_{c}^{0}$
resonances.

The present manuscript is organized as follows. In Section \ref{sec:frame},
we present our theoretical framework, which includes explicit
forms of the employed hyperfine interaction between quarks and the
wave functions of the studied pentaquark configurations.
Numerical results for spectrum
of the studied pentaquark configurations in our model are
shown in Section \ref{sec:result}. Finally, Section \ref{sec:end}
contains a brief conclusion.

\section{Framework}
\label{sec:frame}

In present work, we study the spectrum of low-lying $sscq\bar{q}$ pentaquark configurations
using the constituent quark model, the three-quark version of which has been explicitly depicted in the
literatures~\cite{Glozman:1995fu,Capstick:2000qj}, and five-quark version is recently developed
in~\cite{Yuan:2012zs}. Accordingly, here we just briefly address the key ingredients of
the model. The Hamiltonian in present model is as follow
\begin{eqnarray}
H&=&H_{o}+H_{hyp}+\sum_{i=1}^{5}m_{i}\,,
\label{ham}
\end{eqnarray}
where $m_{i}$ denotes the constituent mass of the
$ith$ quark, $H_{hyp}$ is the
hyperfine interaction between quarks, which is often
treated as perturbation, and $H_{o}$ the Hamiltonian
concerning orbital motions of the quarks which
should contain the kinetic term and
confining potential of the quarks.

Before discussing the Hamiltonian explicitly, now we first concentrate on the wave function for a five-quark system.
Hereafter we label the wave function using the Young tableaux. In the studied five-quark system,
 the color wave function for the four-quark system must be $[211]_{C}$,
which can combine with the color wave function of the antiquark to form a color singlet.
On the other hand, we here take all the quarks and antiquark in the studied pentaquark state to be in their ground
states, namely, the orbital wave function for the four-quark subsystem is the
completely symmetric $[4]_{X}$. Therefore, the flavor-spin wave function of the four-quark subsystem in the studied
five-quark configuration must be $[31]_{FS}$, to form a totally antisymmetric flavor-spin-color-orbital wave function
of the four Fermions system.

The corresponding flavor wave functions of the four-quark
subsystem can be $[4]_{F}$, $[31]_{F}$, $[22]_{F}$ or $[211]_{F}$, and the spin states can be
$[4]_{S}$, $[31]_{S}$ or $[22]_{S}$. One should notice that here we treat the quarks
 with four different flavor as identical particles in the wave functions of the studied
 pentaquark states, and the flavor symmetry breaking effect
 will be included in the Hamiltonian of the five-quark system, in another word, 
 the flavor symmetry breaking effect is treated as perturbation, just analogous to the approach
 employing in~\cite{Glozman:1995fu,Glozman:1995xy}. Consequently, there are $7$ resulting pentaquark states, more comprehensive
analysis for the nucleon-like pentaquark states can be found in Ref.~\cite{Helminen:2000jb},
one can also obtain the same results by analyzing the decomposition of the flavor-spin state $[31]_{FS}$ in the $S_{4}$
permutation group~\cite{chen}.
As we can see in~\cite{Helminen:2000jb}, the two highest states are $300-400$~MeV higher
than the lowest states, even $100-200$~MeV higher than the fifth state. In addition, 
for the sake of the completely symmetric flavor (or spin) configuration, mixing 
between these two highest states and the other five ones is small. 
Therefore, tentatively, here we consider the five lower states which are listed in
Table~\ref{con}. And the general wave function for these states can be written as follow
\begin{eqnarray}
&\psi_{t,s}^{(i)} =\sum_{a,b,c}\sum_{Y,y,T_z,t_z}\sum_{S_z,s_z}
C^{[1^4]}_{[31]_a[211]_a} C^{[31]_a}_{[F^{(i)}]_b [S^{(i)}]_c}
[F^{(i)}]_{b,Y,T_z} [S^{(i)}]_{c,S_z}
[211;C]_a\nonumber\\
&(Y,T,T_z,y,\bar t,t_z|-2,0,t)
(S,S_z,1/2,s_z|S_{\Omega_{c}^{0}},s)\bar\chi_{y,t_z}\bar\xi_{s_z}\varphi_{[5]}\, .
\label{wfc}
\end{eqnarray}
Here $i$ is the number of the $sscq\bar q$ configuration in Table
\ref{con}, $\bar\chi_{y,t_z}$ and $\bar\xi_{s_z}$ represent the
isospinor and the spinor of the antiquark respectively, and
$\varphi_{[5]}$ represents the completely symmetrical orbital wave
function. The first summation involves The symbols
$C^{[.]}_{[..][...]}$, which are $S_4$ Clebsch-Gordan coefficients
for the indicated color ($[211]$), flavor-spin ($[31]$) and flavor
($[F]$) and spin ($[S]$) wave functions of the $sscq$ system. The
second summation runs over the flavor indices in the $SU(4)$
Clebsch-Gordan coefficient and the third over the
spin indices in the standard $SU(2)$ Clebsch-Gordan coefficient. In
the case of the spin configuration $[22]$ the total spin of the
$sscq$ system vanishes, so that $S=S_z=0$.

%%%%%%%%%%%%%%%%%%%%%%%%%%%%%%%%%%%%%%%%%%%%%%%%%%%%%%%%%%%%%%%%%%%%%%%%%%%%%%%%%%%%%%%%%%%%%%%%%%%
%                            Table I                                                              %
%%%%%%%%%%%%%%%%%%%%%%%%%%%%%%%%%%%%%%%%%%%%%%%%%%%%%%%%%%%%%%%%%%%%%%%%%%%%%%%%%%%%%%%%%%%%%%%%%%%
\begin{table*}[ht]
\caption{The studied $sscq\bar{q}$ configurations.
\label{con}}
\vspace{0.3cm}
\renewcommand\tabcolsep{0.49cm}
\renewcommand{\arraystretch}{1.0}
\begin{tabular}{cc}
\hline\hline

$N_{con}$     &  States    \\

\hline

1  &$sscq([4]_{X}[211]_{C}[31]_{FS}[211]_{F}[22]_{S})\otimes \bar{q}$ \\

2  &$sscq([4]_{X}[211]_{C}[31]_{FS}[211]_{F}[31]_{S})\otimes \bar{q}$ \\

3 &  $sscq([4]_{X}[211]_{C}[31]_{FS}[22]_{F}[31]_{S})\otimes \bar{q}$ \\

4&  $sscq([4]_{X}[211]_{C}[31]_{FS}[31]_{F}[22]_{S})\otimes \bar{q}$ \\

5&   $sscq([4]_{X}[211]_{C}[31]_{FS}[31]_{F}[31]_{S})\otimes \bar{q}$ \\

\hline\hline
\end{tabular}
\end{table*}
%%%%%%%%%%%%%%%%%%%%%%%%%%%%%%%%%%%%%%%%%%%%%%%%%%%%%%%%%%%%%%%%%%%%%%%%%%%%%%%%%%%%%%%%%%%%%%%%%%%
%%%%%%%%%%%%%%%%%%%%%%%%%%%%%%%%%%%%%%%%%%%%%%%%%%%%%%%%%%%%%%%%%%%%%%%%%%%%%%%%%%%%%%%%%%%%%%%%%%%

Now we come back to the Hamiltonian of the studied five-quark system.
For the sake of the existence of a charm quark in the present case, we have to
take into account the corrections from $SU(4)$ symmetry breaking. A tentative
way is to consider the corrections by introduction of the following flavor-dependent
Hamiltonian~\cite{Glozman:1995xy}
\begin{equation}
H_{o}^{\prime}=-\sum_{i=1}^{4}\frac{m_{c}-m}{2m}
\frac{\vec{p}_{i}^{2}}{m_{c}}
\delta_{ic}\,,
\end{equation}
where $\delta_{ic}$ is a flavor dependent operator with eigenvalue $1$ for charm quark and $0$
for other quarks. After some explicit calculation, we can find that the matrix elements of the above
Hamiltonian between all the five states listed in Table~\ref{con} are the
same one.

Accordingly, one may notice that if we neglect the hyperfine interaction between quarks in
the five-quark system studied here, then the five states should be degenerate. The degenerate
energy depends on the constituent quark masses, and explicit quark confinement model. For instance,
if one takes the quark confinement model to be the oscillator model, namely, the orbital Hamiltonian
in Eq.~(\ref{ham}) can be written as
\begin{equation}
H_{o}=\sum_{i=1}^5 {\vec{p}_i^2\over 2 m_{i}}-\sum_{i<j}^5
\frac{3}{8}\lambda_i^C\cdot\lambda_j^C\left[C(\vec{r}_i-\vec{r}_j)^2+V_0\right],,
\label{hop}
\end{equation}
then the degenerate energy $E_{0}$ should be
\begin{equation}
E_{0}=2m+2m_{s}+m_{c}+6\sqrt{5C/m}+5V_{0}+\langle H_{o}^{\prime}\rangle\\,
\end{equation}
where $m$, $m_{s}$ and $m_{c}$ denote the constituent masses of the light, strange and charm quarks.
To avoid involving too much parameters, here we take the degenerate energy to be
a free parameter $E_{0}$, without introducing a explicit quark confinement model.

Finally, to get the mass splitting of the studied five pentaquark configurations, we take the $SU(4)$ broken
form of the hyperfine interaction mediated by Goldstone boson exchange as in~\cite{Glozman:1995fu,Yuan:2012zs}
\begin{equation}
H_{hyp}^{GBE}=-\sum_{i,j}^{4}C_{i,j}^{M}\vec\lambda_i^{F}\cdot
\vec\lambda_j^{F} \vec\sigma_i\cdot\vec\sigma_j, \label{fs}
\end{equation}
where $\vec{\lambda}_i^F$ are the
Gell-Mann matrices in flavor space, and $C_{i,j}^{M}$ a
flavor dependent operator for strength of a meson $M$ exchange between
the $ith$ and $jth$ quarks. Notice that in the GBE model,
hyperfine interaction between quark and antiquark
is assumed to be automatically included in the GBE interaction,
so the spin-spin interaction $H_{hyp}^{GBE}$ in Eq.~(\ref{fs}) is
restricted to the four-quark subsystem.

Explicit calculations of the matrix elements of $H_{hyp}^{GBE}$ in~(\ref{fs}) between the five
pentaquark configurations listed in Table~\ref{con} lead to the following results:
\begin{eqnarray}
\mathcal{E}_{K}=
-C^{K}\pmatrix{   7.5    &   0           &   0        &  -3.5   &     0 \cr
            0         &   4.5      &  6.1     &   0      &     1.4\cr
            0         &   6.1       &  4.0   &   0      &   -0.7 \cr
           -3.5      &    0          &   0        & -3.2   &     0   \cr
            0         &   1.4        &   -0.7     &   0      &    2.2 \cr
           }\,,
\label{kex}
\end{eqnarray}
\begin{eqnarray}
\mathcal{E}_{s\bar{s}}=
-C^{s\bar{s}}\pmatrix{   0    &   0           &   0        &  1.5   &     0 \cr
            0         &   1.5      &  0     &   0      &     0\cr
            0         &   0       &  0.5   &   0      &   1.4 \cr
           1.5      &    0          &   0        & 0   &     0   \cr
            0         &   0       &   1.4     &   0      &    -0.5 \cr
           }\,,
\label{sex}
\end{eqnarray}
\begin{eqnarray}
\mathcal{E}_{D}=
-C^{D}\pmatrix{   0    &   0           &   0        &  -2.5   &     0 \cr
            0         &   2.5      &  0     &   0      &     2.9\cr
            0         &   0       &  0.5   &   0      &   -1.4 \cr
           -2.5      &    0          &   0        & 2.7   &     0   \cr
            0         &   2.9        &   -1.4     &   0      &   -0.2 \cr
           }\,,
\label{dex}
\end{eqnarray}
\begin{eqnarray}
\mathcal{E}_{D_{s}}=
-C^{D_{s}}\pmatrix{   7.5    &   0           &   0        &  4.5   &     0 \cr
            0         &   4.5      &  -6.1     &   0      &     -4.3\cr
            0         &   -6.1       &  4.0   &   0      &   0.7 \cr
           4.5      &    0          &   0        & 7.5   &     0   \cr
            0         &   -4.3        &   0.7     &   0      &    3.5 \cr
           }\,,
\label{dsex}
\end{eqnarray}
Where $\mathcal{E}_{M}$ and $C^{M}$ are the hyperfine interaction energy mediated by $M$ meson
and the coupling strength constant for $M$ meson exchange, respectively.

\section{Numerical results}
\label{sec:result}

Before moving to our numerical results, we have to discuss the parameters in present model explicitly.
as shown in Sec. \ref{sec:frame}, the parameters are the degenerate energy $E_{0}$,
and the coupling strength constants in the Goldstone boson exchange
model. Values for the latter have been discussed in~\cite{Yuan:2012zs} for the light five-quark system
and in~\cite{Glozman:1995fu,Glozman:1995xy} for three-quark system. Generally, the empirical values
for the coupling strength constants in present case should be different from the ones given
in~\cite{Yuan:2012zs,Glozman:1995fu,Glozman:1995xy}, because of the existence of the charm quark
in the five-quark system.

In any case, tentatively, if one takes the empirical values for the coupling strength constants in the literature~\cite{Yuan:2012zs,Glozman:1995fu,Glozman:1995xy},
namely, $C^{K}=15.5$~MeV, $C^{s\bar{s}}=11.5$~MeV, $C^{D}=6.5$~MeV and $C^{D_{s}}=6.5$~MeV, and treats
the degenerate energy $E_{0}$ as free parameter to fit most of the observed energies of $\Omega_{c}^{0}$ resonances,
which yields $E_{0}=3132$~MeV, one can get the energy matrix of the five pentaquark configurations listed
in Table~\ref{con}:
\begin{eqnarray}
E=
\pmatrix{   2967    &   0           &   0        &  24   &     0 \cr
            0         &   2999      &  -55     &   0      &     -13\cr
            0         &   -55       &  3035   &   0      &  -1 \cr
           24      &    0          &   0        & 3115   &     0   \cr
            0         &   -13        &   -1     &   0      &    3082 \cr
           }\,,
\label{enq}
\end{eqnarray}
in unit of MeV. As we can see, there are several nonnegligible nondiagonal matrix elements,
which must lead to mixing of the configurations given in Table~\ref{con}. Diagonalization of the $E$ matrix~(\ref{enq}) leads to physical masses and the configuration
mixing coefficients given in Table~\ref{enf}.

One may notice that value for the free parameter $E_{0}$ in present work is
about $\sim1000$ MeV higher than the one obtained by fitting the data for sea content in nucleon~\cite{An:2012kj}. As we know, the
constituent quark mass of charm quark should be $\sim1200-1300$~MeV higher than that of light quark, while
the studied pentaquark configurations in~\cite{An:2012kj} are $1/2^{+}$ states, namely, those configurations
are the first orbitally excited pentaquark states, which should raise up $E_{0}$ by about $200-300$~MeV. Accordingly, the value
for $E_{0}$ obtained in present work is reasonable.
%%%%%%%%%%%%%%%%%%%%%%%%%%%%%%%%%%%%%%%%%%%%%%%%%%%%%%%%%%%%%%%%%%%%%%%%%%%%%%%%%%%%%%%%%%%%%%%%%%%
%                            Table II                                                              %
%%%%%%%%%%%%%%%%%%%%%%%%%%%%%%%%%%%%%%%%%%%%%%%%%%%%%%%%%%%%%%%%%%%%%%%%%%%%%%%%%%%%%%%%%%%%%%%%%%%
\begin{table*}[ht]
\caption{The masses of the studied $sscq\bar{q}$ pentaquark states and configuration mixing
coefficiens. Numbers in the first row are the obtained energies for the five pentaquark
states in unit of MeV, and numbers in columns 2-6 are the corresponding mixing coefficients
for the configurations in Table~\ref{con}.
\label{enf}}
\vspace{0.3cm}
\renewcommand\tabcolsep{0.49cm}
\renewcommand{\arraystretch}{0.8}
\begin{tabular}{cccccc}
\hline\hline

              &  2958  &  2963   &  3071  &  3088  &  3119  \\

\hline

$|1\rangle$   &  0     &  0.988  &  0     &   0    &   0.157\\

$|2\rangle$   &  0.808 &  0      &  0.471  &   0.354 &  0  \\

$|3\rangle$   &  0.583 &  0      &  -0.728 &   -0.361 & 0 \\

$|4\rangle$   &  0     &  -0.157      &  0      &  0       & 0.988 \\

$|5\rangle$   &  0.088  & 0      &  0.498       &  -0.863  & 0 \\

\hline\hline
\end{tabular}
\end{table*}
%%%%%%%%%%%%%%%%%%%%%%%%%%%%%%%%%%%%%%%%%%%%%%%%%%%%%%%%%%%%%%%%%%%%%%%%%%%%%%%%%%%%%%%%%%%%%%%%%%%
%%%%%%%%%%%%%%%%%%%%%%%%%%%%%%%%%%%%%%%%%%%%%%%%%%%%%%%%%%%%%%%%%%%%%%%%%%%%%%%%%%%%%%%%%%%%%%%%%%%

As we can see in Table~\ref{enf}, the numerical results fit LHCb data very well, with the largest
deviation at $\sim3\%$. As we have expected, mixing between the configurations $2$, $3$ and $5$ in Table~\ref{con} are
strong, which corrects the diagonal energies in~(\ref{enq}) at $\sim1\%$.

As we have discussed at the beginning of this section, the coupling strength constants for Goldstone boson
exchange in present work may be different from those in the literatures, so here we try to deviate the values
we have taken by $\pm10\%$. The resulting numerical results are given in Table~\ref{enp} compared to the experimental
data and results obtained in other approaches. As shown in Table~\ref{enp}, masses of the three higher states
are not very sensitive to the coupling strength constants. Accordingly, we may conclude that these three pentaquark
configurations can form sizable components in the observed $\Omega_{c}(3066)^{0}$, $\Omega_{c}(3090)^{0}$ and
$\Omega_{c}(3119)^{0}$ states, and the quantum numbers for the observed three states may be $1/2^{-}$.

%%%%%%%%%%%%%%%%%%%%%%%%%%%%%%%%%%%%%%%%%%%%%%%%%%%%%%%%%%%%%%%%%%%%%%%%%%%%%%%%%%%%%%%%%%%%%%%%%%%
%                            Table III                                                              %
%%%%%%%%%%%%%%%%%%%%%%%%%%%%%%%%%%%%%%%%%%%%%%%%%%%%%%%%%%%%%%%%%%%%%%%%%%%%%%%%%%%%%%%%%%%%%%%%%%%

\begin{table*}[ht]
\caption{The masses of the studied $sscq\bar{q}$ pentaquark states compared to the experimental data
and results in other approaches. All the numbers are in unit of MeV.
\label{enp}}
\vspace{0.3cm}
\renewcommand\tabcolsep{0.1cm}
\renewcommand{\arraystretch}{0.1}
\begin{tabular}{cccccccc}
\hline\hline

  Approach       & && $M_{\Omega_{c}^{0}}$ &&&  Ref \\

  \hline

  Data           &  $3000$       &  $3050$        &  $3066$      &  $3090$      & $3119$       & \cite{Aaij:2017nav}\\

  CQM            &  $2958\pm18$  &  $2963\pm17$   &  $3071\pm6$  &  $3088\pm4$  &  $3119\pm1$ &  Present \\

 QCD Sum Rule    &               &                &              &  $3066\pm138$  &  $3119\pm114$     & \cite{Agaev:2017jyt}              \\

 QCD Sum Rule    &               &                &      $3100~?$        &   $3120~?$   &  $3100~?$         &\cite{Chen:2017sci}          \\

 QCD Sum Rule    &               &  $3050\pm110$  &      $3060\pm110$    &    $3060\pm100$  & $3110\pm100$  & \cite{Wang:2017zjw}          \\

 Quark-Diquark  &  $2987$       &  $3005$        &  $3077$      &  $3095$      &  $3227$     &  \cite{Wang:2017vnc}  \\

    $\chi$QM            &               &                 &              &                    &  $3156$           &\cite{Yang:2017rpg}\\

    Quark-Diquark &  $2975$      &  $3057$       &      $3066$           &    $3063$          &  $3120$       &   \cite{Chen:2017gnu} \\

\hline\hline
\end{tabular}
\end{table*}

%%%%%%%%%%%%%%%%%%%%%%%%%%%%%%%%%%%%%%%%%%%%%%%%%%%%%%%%%%%%%%%%%%%%%%%%%%%%%%%%%%%%%%%%%%%%%%%%%%%
%%%%%%%%%%%%%%%%%%%%%%%%%%%%%%%%%%%%%%%%%%%%%%%%%%%%%%%%%%%%%%%%%%%%%%%%%%%%%%%%%%%%%%%%%%%%%%%%%%%

Our numerical results for the two lowest states have sizable deviation from the experimental data,
especially the second one. While one may also expect that the next-to-lowest state in our model may
correspond for the observed $\Omega_{c}(3000)^{0}$, since once the values of coupling strength constants
for Goldstone boson exchange are taken to be $0.9$ times of the ones given in~~\cite{Yuan:2012zs,Glozman:1995fu,Glozman:1995xy},
the resulting energy for the next-to-lowest state is $2980$~MeV, while the three higher states
almost keep as what they are.

In addition, one has to notice that the third and fourth obtained states in our model (may correspond to the
observed $\Omega_{c}(3066)^{0}$ and $\Omega_{c}(3090)^{0}$ resonances),
which are admixture of the configurations $|2\rangle$, $3\rangle$ and $|5\rangle$ in Table~\ref{con}, may also have spin quantum
number $J=3/2$. This is because of that spin wave functions for the four-quark subsystems of these three configurations
 are $[31]_{S}$, namely, $S_{4q}=1$,
which could result in $S_{5q}=1/2$ or $S_{5q}=3/2$ when combine with the spin of the antiquark. While in present model,
the energies of spin $1/2$ and $3/2$ states are the same.

In Refs.~\cite{Agaev:2017jyt,Chen:2017sci,Wang:2017zjw,Aliev:2017led}, $\Omega_{c}^{0}$ states are investigated using
the QCD sum rule approach. While the obtained quantum numbers of $\Omega_{c}^{0}$ states are very different.
In~\cite{Agaev:2017jyt},  $\Omega_{c}(3066)^{0}$ and $\Omega_{c}(3119)^{0}$ are interpreted to be
$(2S, 1/2^{+})$ and $(2S, 3/2^{+})$ charmed baryons. The $\Omega_{c}(3000)^{0}$, $\Omega_{c}(3050)^{0}$ and $\Omega_{c}(3066)^{0}$
are assigned to be $1/2^{-}$ baryon states. In~\cite{Wang:2017zjw,Aliev:2017led},
all of the $\Omega_{c}^{0}$ resonances have negative parity, but with different spin quantum numbers.

The $\Omega_{c}^{0}$ resonances have also been investigated by using various of quark
models~\cite{Karliner:2017kfm,Wang:2017vnc,Yang:2017rpg,Wang:2017hej,Chen:2017gnu} since
the observation by LHCb collaboration. In~\cite{Karliner:2017kfm,Wang:2017vnc,Chen:2017gnu},
the $\Omega_{c}^{0}$ states are interpreted as system with one diquark and one charm quark,
and the parities for all the five observed states are assigned to be negative, but with different spin quantum numbers.
while in~\cite{Wang:2017hej}, the parity quantum number of the highest $\Omega_{c}^{0}$ state
is assigned to be positive but the other four states are proposed to have negative parity. In~\cite{Yang:2017rpg},
the $\Omega_{c}^{0}$ states with pentaquark structure are first investigated using the chiral
quark model.

There are several other works in which the quantum numbers of the five observed $\Omega_{c}^{0}$ baryons
are investigated~\cite{Padmanath:2017lng,Cheng:2017ove}. Two of the five states are identified to have
positive parity, while the other three have negative parity in~\cite{Cheng:2017ove}. The lattice QCD
calculations~\cite{Padmanath:2017lng} indicate that $\Omega_{c}(3000)^{0}$ and $\Omega_{c}(3050)^{0}$
may have quantum numbers $1/2^{-}$, and $\Omega_{c}(3066)^{0}$ and $\Omega_{c}(3090)^{0}$ may have $J^{P}=3/2^{-}$,
whereas $\Omega_{c}(3119)^{0}$ may be a $5/2^-$ state.

\section{Summary}
\label{sec:end}

In present work, we investigate the spectrum of several low-lying $sscq\bar{q}$ pentaquark configurations with $I=0$
employing the constituent quark model, within which the hyperfine interaction between quarks is taken
to be mediated by Goldstone boson exchange. Mixing of different pentaquark configurations
caused by hyperfine interaction is considered. And because of the existence of the charm quark, the
$SU(4)$ flavor symmetry breaking effects from both the constituent masses for different quark and coupling
strength constants for different meson exchanges are also taken into account.

In fact, it's not so convenient for us to compare the numerical results to the experimental data,
since there is no explicit experimental information on quantum numbers of the five newly observed $\Omega_{c}^{0}$ states,
and the various of different theoretical investigations give very different predictions.
In any case, tentatively, with reasonable values for the model parameters,
our numerical results show that masses of three obtained $sscq\bar{q}$ states are very close
to the newly observed $\Omega_{c}(3066)^{0}$,  $\Omega_{c}(3090)^{0}$ and $\Omega_{c}(3119)^{0}$,
which may indicate that those obtained $sscq\bar{q}$ configurations may form sizable components in
these $\Omega_{c}^{0}$ baryons. In addition, another obtained $sscq\bar{q}$ state lies at $\sim2980$~MeV, it's
also very close to the mass of the observed $\Omega_{c}(3000)^{0}$, with the deviation less than $1\%$.

To conclude, the present estimation on the masses of low-lying $sscq\bar{q}$ configurations
show that the pentaquark configurations may take sizable probabilities in the observed $\Omega_{c}^{0}$
baryons. Considering the quantum numbers, $\Omega_{c}(3000)^{0}$ and $\Omega_{c}(3119)^{0}$ may be $1/2^{-}$ states, while
$\Omega_{c}(3066)^{0}$ and $\Omega_{c}(3090)^{0}$ may have the spin-parity quantum number
$1/2^{-}$ or $3/2^{-}$.

\begin{acknowledgments}
We are indebted to Prof. B. S. Zou for the suggestion to consider the $\Omega_{c}^{0}$
resonances and very useful discussions.
This work is partly supported by the National Natural Science Foundation of China under Grant
Nos. 11675131 and 11645002, the Chongqing Natural Science Foundation under Grant No. cstc2015jcyjA00032, and the
Fundamental Research Funds for the Central Universities under Grant No. SWU115020.
\end{acknowledgments}

%%%%%%%%%%%%%%%%%%%%%%%%%%%%%%%%%%%%%%%%%%%%%%%%%%%%%%%%%%%%%%%%%%%%%%%%%%%%%%%%%%%%%%%%%%%%%%%%%%%%%%%%%%%
%                                 References                                                              %
%%%%%%%%%%%%%%%%%%%%%%%%%%%%%%%%%%%%%%%%%%%%%%%%%%%%%%%%%%%%%%%%%%%%%%%%%%%%%%%%%%%%%%%%%%%%%%%%%%%%%%%%%%%

%%%%%%%%%%%%%%%%%%%%%%%%%%%%%%%%%%%%%%%%%%%%%%%%%%%%%%%%%%%%%%%%%%%%%%%%%%%%%%%%%%%%%%%%%%%%%%%%%%%%%%%%%%%%%%%%%%%%%%%%%

\end{document}